\documentstyle[epsf,aps]{revtex}
\newcommand{\be}{\begin{equation}}
\newcommand{\ee}{\end{equation}}
\newcommand{\bc}{\begin{center}}
\newcommand{\ec}{\end{center}}
\begin{document}

\title{Low-lying excitations and thermodynamics of an antiferromagnetic
Heisenberg fractal system of a dimension between one and two}

\author{Andreas Voigt
\footnote{\footnotesize Tel. +49-391-67-12473; fax.
+49-391-67-11217; e-mail: Andreas.Voigt@Physik.Uni-Magdeburg.de},
Johannes Richter and Piotr Tomczak$^1$}

\address{Institut f\"ur Theoretische Physik\\
Otto-von-Guericke-Universit\"at Magdeburg, Postfach 4120, 39106
Magdeburg, Germany\\
$^1$ Institute of  Molecular Physics, Polish
Academy of Sciences,\\ Smulochowskiego 17, PL-60-179 Poznan, Poland}

\date{\today}

\maketitle

\begin{abstract}
We investigate a frustrated Heisenberg spin-$1\over2$ antiferromagnet on a
fractal lattice of dimension d=ln3/ln2 (Sierpi\'{n}ski gasket).
Calculations were performed using (a) exact diagonalization of all
eigenstates and eigenvectors for systems up to N=15 and (b) the
Decoupled-Cell Quantum-Monte-Carlo method for systems up to $N=366$. We
present the low-lying spectrum and the specific heat. The specific heat
shows a second maximum in the low-temperature region. This behavior is
similar to the behavior of the quantum Heisenberg antiferromagnet on a
kagom\'{e} lattice and suggests a disordered ground state and a spin gap
in the considered system. \end{abstract}

{\bf Keywords:}
Heisenberg antiferromagnet, Fractals, specific heat -- low temperature

{\bf Introduction}.
The question for the ground state ordering and the low-temperature
thermodynamics of quantum antiferromagnets in one or two dimensions
receives much interest. In $d=1$ there is no long-range order (LRO) in
the ground state. For $d=2$ this question is not generally answered.
Various features may influence the magnetic ground state ordering:
frustration, low coordination number, topological disorder. For several
lattices (e.g. square, honeycomb, triangular) there is evidence for a
long-range ordered ground state, indicating some preference for an
ordered ground state in $d=2$. Only for the kagom\'{e} lattice the
ground state may be disordered. Recently \cite{Tomczak:1996,TFR:1996},
the question has been discussed 'What happens in a quantum
antiferromagnet with a dimension between one and two?'. In
\cite{TFR:1996} two of of us (P.T. and J.R.) investigated the ground
state of the spin-$1\over2$ system on the Sierpi\'{n}ski gasket
(fractal dimension $d=\ln3/\ln2)$ by exact diagonalization
and variational Monte-Carlo method. We found rather strong arguments
that the spin system on the Sierpi\'{n}ski gasket has no long range
order and that its ground state may correspond to a spin liquid.

Additionally a quantum decimation technique was used in
\cite{Tomczak:1996} by one of us (P.T.) to investigate finite temperature
properties of that system. It was argued, that the Sierpi\'{n}ski gasket
has some similarity with the kagom\'{e} lattice.

In this paper we want to analyze the low-lying spectrum and finite
temperature properties of the Sierpi\'{n}ski gasket by alternative
approaches, i.e. quantum Monte Carlo and full diagonalization of small
systems.

{\bf The model}.
We consider the following Hamiltonian
\be
H =  J \sum_{<i,j>} \vec S_i \vec S_j
\ee
where $\vec S_i$ is the spin-$1\over2$ operator at the site $i$ and the
summation $<i,j>$ runs over all nearest neighbors on a Sierpi\'{n}ski
fractal lattice. (A system with 366 sites is presented in Fig. 1). The
strength of the antiferromagnetic interactions is $J=1$. The classical
ground state of this system is exactly known: The spins belonging to
three interpenetrating, equivalent sublattices form a $120^\circ$
structure.

For the quantum system we have calculated all eigenvalues and
eigenstates for the systems with $N=6$ and $N=15$. (The next size is
$N=42$ and goes beyond all computer facilities available at present).
To consider larger systems we have, alternatively, used a decoupled-Cell
Quantum Monte-Carlo approach.

{\bf Results}.
In Fig.2 we present the low-lying spectrum  belonging to different
quantum numbers $S$ of the total spin ${\bf S}^2$ for the system with
$N=15$. According to \cite{Bernu:1992}  a symmetry broken N\'{e}el like
LRO can be realized in the thermodynamic limit ($N \rightarrow \infty$)
if the lowest levels could be described by an effective Hamiltonian $
H_{eff} = E_0 + (K/N){\bf S}^2$, i.e. the lowest $E_0(S)$ should be
linearly dependent on $S(S+1)$. From Fig.2 it is obvious that for the
Sierpi\'{n}ski gasket there is a significant deviation from that linear
behavior (cf. straight line between $E_0(S=1/2)$ and $E_0(S=7/2)$.
Notice, that this deviation is even much larger than the corresponding
deviation for a linear chain of $16$ sites. There is a second point of
interest concerning ground state ordering: For lattices with
antiferromagnetic zero-temperature LRO the ground state has minimum
total spin (i.e. $S_{GS}=0$ for $N$ even or $S_{GS}=1/2$ for $N$ odd)
\cite{Lieb:1962,Richter:1992} and the first excitation has a total spin
$S=S_{GS}+1$. In accordance with the kagom\'{e} lattice
\cite{Lecheminant:1997,Everts:1997} we find here (cf. Fig.2) (i) a
twofold non-trivial degenerated ground state and (ii) 10 additional
$S=1/2$ states between the ground state and the lowest $S=3/2$ state.
Having in mind the above discussion and  the variational calculation of
the spin correlations in \cite{TFR:1996} we argue that for the
Sierpi\'{n}ski gasket a symmetry broken N\'{e}el like ordering is not
favored.

Another interesting feature of the lowest energies of these system is
the existence of two gaps between the lowest eigenvalues in the subspace
$S=1/2$. Numerical calculations show that only these two gaps are
necessary in order to have two additional peaks in the low-temperature
region of the specific heat.

In Fig.3 we present the specific heat as a function of the temperature
$kT/J$ from the exact diagonalization data for $N=6$ and $N=15$ and from
decoupled-cell Quantum-Monte-Carlo calculations for $N=366$. This
Monte-Carlo approach was introduced by Homma {\it et al}
\cite{Homma:1984} and later applied successfully to several quantum spin
systems (see e.g. \cite{DCM}). For this method the full diagonalization
of small clusters is necessary in order to calculate the transition
probability in the Monte-Carlo process. For the calculations we have
used clusters with $N=6,15$. In Fig.3 we present the data with the
$N=15$ cluster. We find one additional peak in the low temperature
region for the system with $N=6$ and as mentioned above even two such
peaks for the system with $N=15$. The Quantum-Monte-Carlo data for the
system with $N=366$ also show one additional peak in the low temperature
region (The symbols are as high as the statistical error of the data).
This behavior in the specific heat corresponds to the results obtained
by the quantum decimation technique and is quite similar to that of a
kagom\'{e} lattice \cite{Zeng:1990} and supports the suggestion of a
disordered ground state.

{\bf Conclusion.}
In general, we argue that the similarity to the kagom\'{e} lattice may
be connected with the lack of LRO in both systems. If magnetic
correlations are short-ranged the local lattice structure plays the
important role. The local environment of both lattices  is quite similar
(four nearest neighbors, triangle as an elementary unit).

{\bf Acknowledgments}.
This work has been supported by the DFG (Ri 615/1-2) and by the KBN
(8T11F 010 08p04).




{\bf Fig.1:} The Sierpi\'{n}ski gasket with $N=366$.


{\bf Fig.2:} The lowest energy eigenvalues of the Sierpi\'{n}ski gasket
with $N=15$ versus $S(S+1)$ ($S$ is the quantum number
of the total spin).


{\bf Fig.3:} The specific heat (per spin) versus temperature.

\end{document}